\documentclass[10pt,oneside,onecolumn]{elsart}

\usepackage[ansinew]{inputenc}
\usepackage[english]{babel}
\usepackage{ifthen,shortvrb}
\usepackage[fleqn]{amsmath}
\usepackage{amsxtra,amsfonts,latexsym,amssymb,euscript}
\usepackage{amstext}
\usepackage{graphicx}
\usepackage{dcolumn}
\usepackage{cite}
\usepackage{subfigure}

\def¡{\hphantom{-}}      % alt-173
\def«{\hphantom{0}}      % alt-174
\def»{\hphantom{.}}      % alt-175

\begin{document}

\begin{frontmatter}
%\journal{}
\title{A theoretical approach to the interaction between buckling and resonance instabilities}
\author{Alberto Carpinteri\corauthref{gz}},
\corauth[gz]{Corresponding author. Tel. +39-011-564-4850 Fax
+39-011-564-4899} \ead{alberto.carpinteri@polito.it}
\author{Marco Paggi}
\address{Politecnico di Torino, Department of Structural and Geotechnical
Engineering, Corso Duca degli Abruzzi 24, 10129 Torino, Italy}

\begin{abstract}
The paper deals with the interaction between buckling and resonance
instabilities of mechanical systems. Taking into account the effect
of geometric nonlinearity in the equations of motion through the
geometric stiffness matrix, the problem is reduced to a generalized
eigenproblem where both the loading multiplier and the natural
frequency of the system are unknown. According to this approach, all
the forms of instabilities intermediate between those of pure
buckling and pure forced resonance can be investigated. Numerous
examples are analyzed, including: discrete mechanical systems with
one to $n$ degrees of freedom, continuous mechanical systems such as
oscillating deflected beams subjected to a compressive axial load,
as well as oscillating beams subjected to lateral-torsional
buckling. A general finite element procedure is also outlined, with
the possibility to apply the proposed approach to any general bi- or
tri-dimensional framed structure. The proposed results provide a new
insight in the interpretation of coupled phenomena such as flutter
instability of long-span or high-rise structures.
\end{abstract}
\end{frontmatter}

\section{Introduction}

Buckling, resonance and flutter are the main forms of instability of
the elastic equilibrium of structural systems. In buckling
instability, by removing the hypothesis of small displacements so
that the deformed structural configuration can be distinguished from
the undeformed one, it is possible to show that the solution of an
elastic problem can represent a condition of stable, neutral or
unstable equilibrium, depending on the magnitude of the applied
load. As is well-known, buckling is usually observed in slender
structural elements subjected to a compressive stress field, such as
columns of buildings, machine shafts, struts of trusses, thin arches
and shells. In some cases, elastic instability can also take place
for special loading and geometrical conditions, as for example in
the lateral torsional buckling of slender beams \cite{aC97}.

The phenomenon of resonance is also particularly important in
structural engineering. It represents a form of dynamic instability,
which occurs when an external periodic frequency matches one of the
natural frequencies of vibration of the mechanical system. In this
case, therefore, structural design deals with the determination of
such dangerous natural frequencies according to modal analysis
\cite{CP75,CP05a,CP05b}.

Finally, the phenomenon of flutter is a form of aeroelastic
instability observed in long-span or high-rise structures subjected
to wind loads, such as towers, tall buildings \cite{hK93} and
suspended \cite{mA98,rS01,rhS96} or cable-stayed bridges
\cite{uS93}. In this case, the instability is attributed to
motion-induced or self-excited forces, which are loads induced or
influenced by the deformation of the structure itself
\cite{SS86,BS91}. Forces originated in this way modify the initial
deformation of the mechanical system which, consequently, leads to
modified forces, and so on. This feed-back mechanism can give rise
to an amplification effect on the initial deformation, leading to
premature failure of the structure. The well-known dramatic Tacoma
Narrows bridge disaster of 1940 is a famous example of this
catastrophic interaction, and it is still very much in the public
eye today. In this field, which is not at present completely
understood, aeroelastic instability is often considered as the
result of the interaction between buckling (static) and resonance
(dynamic) instabilities. However, only a few theoretical
formulations have been proposed for modelling aerodynamic forces
and, in most investigations, empirical models are set up in which
the parameters related to the fluid-structure interaction are
established by experiments \cite{rhS94}.

In the present contribution, we deal with the phenomenon of
interaction between buckling and resonance instabilities. A
state-of-the-art survey of the existing Literature shows that this
problem has been mainly addressed in the field of multi-parameter
stability theory \cite{amL92,vvB63,vvB64,apS91,SM03,kH86,kH02,EA05},
where the conditions for stability of a mechanical system are
studied with reference to a perturbation of the problem parameters.
In this framework, instability domains of some continuous
oscillatory systems subjected to buckling loads were provided
\cite{SM03,kH86}. However, to make the problem analytically
treatable, the analysis was mainly limited to certain continuous
mechanical systems such as deflected beams and beams experiencing
lateral torsional buckling. More importantly, the concept of
geometric stiffness matrix, typical of structural engineering
approaches, was totally neglected.

On the other hand, in the field of bridge engineering, the influence
of the geometric nonlinearity is usually taken into account by
including the contribution of the self-induced aerodynamic forces to
the applied loads (see e.g. \cite{ST71,rhS78,uS93} and \cite{HP02}
for a detailed overview of the mathematical methods). Such
nonconservative aerodynamic forces are put in relationship with the
displacements and the velocities of the points of the mechanical
system according to the so-called \emph{flutter derivatives} that
have to be experimentally determined in the laboratories
\cite{rhS81}. In these approaches, the onset of flutter instability
corresponds to the condition of vanishing structural dumping.
Therefore, structural damping seems to play a fundamental role,
although it is practically impossible to be analytically evaluated
but only experimentally estimated \cite{ST71}.

From the mathematical point of view, it is important to note that
pure buckling, pure resonance, and also flutter instabilities are
usually mathematically treated as eigenvalue problems. In this
paper, we propose a mathematical theory for the analysis of buckling
and resonance interaction, with the possibility to give an insight
onto the mechanisms leading to flutter instability. In the
mathematical treatment, a special focus will be given to the role
played by the geometric stiffness matrix, which contributes to the
reduction of the global elastic stiffness due to the effect of the
geometric nonlinearity. This represents a novelty of our approach
with respect to the models available in the Literature. As it will
be shown in the sequel, the use of the geometric stiffness matrix
may provide the proper link between multi-parameter stability
theory, typical of rational mechanics, and the bridge engineering
approach, typical of bridge engineers.

According to this formulation, we will demonstrate that the
interaction between buckling and resonance leads to a generalized
eigenvalue problem where both the buckling loads and the natural
frequencies of the system are unknown and represent the eigenvalues.
This approach will permit to inspect all the forms of structural
instability intermediate between pure buckling and pure resonance.
These limit cases are instead observed either when the dynamics of
the system is neglected, or when the external buckling forces are
equal to zero.

The effectiveness of the proposed methodology will be demonstrated
with respect to not only discrete mechanical systems with one to $n$
degrees of freedom, but also for continuous mechanical systems such
as oscillating deflected beams and beams showing lateral-torsional
coupled deformations. Finally, a general procedure is established in
the framework of the finite element method.

\section{Discrete mechanical systems}

\subsection{Discrete mechanical systems with one degree of freedom}

\begin{figure}
\begin{center}
\includegraphics*[width=.3\textwidth,angle=-90]{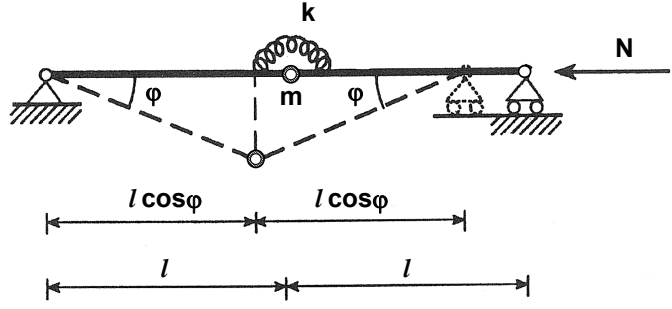}
\caption{Scheme of the first one-degree of freedom system analyzed.}
\label{fig1}
\end{center}
\end{figure}

Let us consider the mechanical system shown in Fig.\ref{fig1},
consisting of two rigid rods connected by an elastic hinge of
rotational rigidity $k$ and constrained at one end by a hinge and at
the other by a roller support. A mass $m$ is placed in
correspondence of the intermediate elastic hinge and the system is
loaded by a horizontal axial force $N$. Considering the absolute
rotation $\varphi$ of the two arms as the generalized coordinate,
the total potential energy, $W$, and the kinetic energy, $T$, of the
whole system are:
\begin{equation}
\begin{aligned}\label{eq1}
W(\varphi)&=\dfrac{1}{2}k(2\varphi)^2-2Nl(1-\cos\varphi),\\
T(\dot{\varphi})&=\dfrac{1}{2}m\left[\dfrac{\mbox{d}}{\mbox{d}t}(l\sin\varphi)\right]^2+\dfrac{1}{2}m\left[\dfrac{\mbox{d}}{\mbox{d}t}(l-l\cos\varphi)\right]^2=\dfrac{1}{2}ml^2\dot{\varphi}^2.
\end{aligned}
\end{equation}

The equation of motion can be determined by writing the Lagrange's
equation:
\begin{equation}\label{eq2}
\dfrac{\partial}{\partial t} \left(\dfrac{\partial
T}{\partial\dot{\varphi}}\right)-\dfrac{\partial
T}{\partial\varphi}=-\dfrac{\partial W}{\partial\varphi}.
\end{equation}
In the present case, this yields:
\begin{equation}\label{eq3}
ml^2\ddot{\varphi}=-4k\varphi+2Nl\sin\varphi,
\end{equation}
which can be suitably linearized in correspondence of $\varphi=0$:
\begin{equation}\label{eq4}
ml^2\ddot{\varphi}=-4k\varphi+2Nl\varphi.
\end{equation}

Looking for the solution to Eq.\eqref{eq4} in the general form
$\varphi=\varphi_0\mbox{e}^{\mbox{i} \omega t}$, where $\omega$
denotes the natural angular frequency of the system, we obtain the
following equation which provides the conditions of equilibrium of
the system:
\begin{equation}\label{eq5}
\left(4k-2Nl-\omega^2 ml^2\right)\varphi_0=0.
\end{equation}
A nontrivial solution to Eq.\eqref{eq5} exists if and only if the
term in brackets is equal to zero. This critical condition
corresponding to the bifurcation of the equilibrium establishes a
one-to-one relationship between the applied axial force, $N$, and
the angular frequency, $\omega$:
\begin{equation}\label{eq6}
N=\dfrac{2k}{l}-\dfrac{ml}{2}\omega^2.
\end{equation}
Moreover, Eq.\eqref{eq6} admits two important limit conditions for,
respectively, $N=0$ and $m=0$. In the former case, Eq.\eqref{eq6}
gives the natural angular frequency of the system according to pure
modal analysis, that is $\omega_1=\sqrt{4k/ml^2}$. In the latter,
the pure critical Eulerian load is obtained, that is $N_1=2k/l$.
Dividing Eq.\eqref{eq6} by $N_1$, we obtain the following
relationship between $N$ and $\omega$ in a nondimensional form:
\begin{equation}\label{eq7}
\left(\dfrac{\omega}{\omega_1}\right)^2+\left(\dfrac{N}{N_1}\right)=1.
\end{equation}

\begin{figure}
\begin{center}
\includegraphics*[width=.32\textwidth,angle=-90]{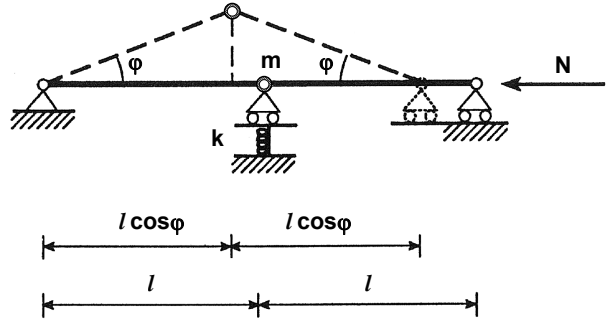}
\caption{Scheme of the second one-degree of freedom system
analyzed.} \label{fig2}
\end{center}
\end{figure}

As a second example, let us consider the mechanical system shown in
Fig.\ref{fig2}, consisting in two rigid rods on three supports, of
which the intermediate one is assumed to be elastically compliant
with rigidity $k$. As in the previous case, a mass $m$ is placed in
correspondence of the intermediate hinge and the system is loaded by
a horizontal axial force $N$. Considering the absolute rotation
$\varphi$ of the two arms as the generalized coordinate, the total
potential energy, $W$, and the kinetic energy, $T$, of the whole
system are:
\begin{equation}
\begin{aligned}\label{eq8}
W(\varphi)&=\dfrac{1}{2}k(l\sin\varphi)^2-2Nl(1-\cos\varphi),\\
T(\dot{\varphi})&=\dfrac{1}{2}ml^2\dot{\varphi}^2.
\end{aligned}
\end{equation}

Following the procedure discussed above, we determine the equation
of motion by employing the Lagrange's equation \eqref{eq2}:
\begin{equation}\label{eq9}
ml^2\ddot{\varphi}=-l\sin\varphi(kl\cos\varphi-2N),
\end{equation}
which can be suitably linearized in correspondence of $\varphi=0$:
\begin{equation}\label{eq10}
ml^2\ddot{\varphi}=-l\varphi(kl-2N).
\end{equation}

Looking for the solution to Eq.\eqref{eq10} in the general form
$\varphi=\varphi_0\mbox{e}^{\mbox{i} \omega t}$, where $\omega$
denotes the natural angular frequency of the system, we obtain the
following condition of equilibrium of the system:
\begin{equation}\label{eq11}
\left(kl^2-2Nl-\omega^2 ml^2\right)\varphi_0=0.
\end{equation}
As in the previous example, by setting equal to zero the term in
brackets, we obtain a one-to-one relationship between the applied
axial force, $N$, and the angular frequency, $\omega$:
\begin{equation}\label{eq12}
N=\dfrac{kl}{2}-\dfrac{ml}{2}\omega^2.
\end{equation}
This equation admits two important limit conditions for,
respectively, $N=0$ and $m=0$. In the former case, Eq.\eqref{eq12}
gives the natural angular frequency of the system according to pure
modal analysis, that is $\omega_1=\sqrt{k/m}$. In the latter, the
pure critical Eulerian load for buckling instability is obtained,
that is $N_1=kl/2$. Dividing Eq.\eqref{eq12} by $N_1$, we obtain the
same relationship between the nondimensional terms $N/N_1$ and
$(\omega/\omega_1)^2$ as in the previous example (see
Eq.\eqref{eq7}).

\begin{figure}
\begin{center}
\includegraphics*[width=.5\textwidth,angle=-90]{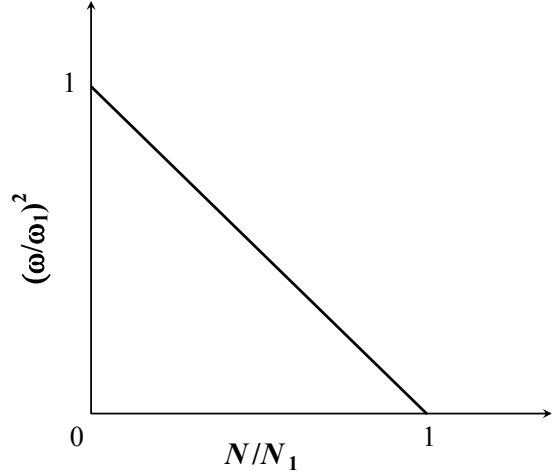}
\caption{Nondimensional frequency vs. nondimensional axial force for
the single degree of freedom systems.} \label{fig3}
\end{center}
\end{figure}

A graphical representation of the condition \eqref{eq7} in
Fig.\ref{fig3} shows that the resonance frequency is a decreasing
function of the compressive axial load. This demonstrates, for the
mechanical systems with a single degree of freedom, that the
condition of bifurcation of the equilibrium can be reached for a
compressive axial force, $N$, lower than the Eulerian buckling load,
$N_1$, provided that the system is subjected to an external
excitation with frequency $\omega$ given by Eq.\eqref{eq7}.
Conversely, failure due to resonance can take place for
$\omega<\omega_1$, provided that the system is loaded by an axial
force $N$ given by Eq.\eqref{eq7}.

Finally, the issue of stability or instability of the mechanical
system in the correspondence of the bifurcation point can be
discussed as in the static case, i.e. by evaluating the higher order
derivatives of the total potential energy $W$.

\subsection{Discrete mechanical systems with $n$ degrees of freedom}

\begin{figure}
\begin{center}
\includegraphics*[width=.25\textwidth,angle=-90]{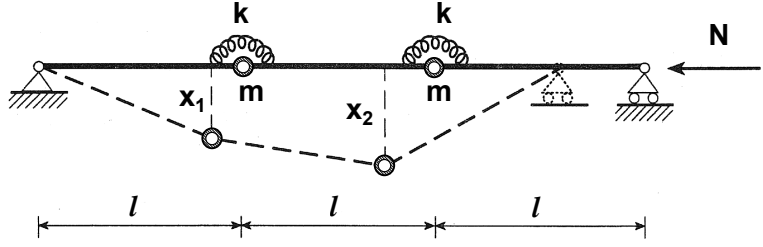}
\caption{Scheme of the two-degrees of freedom system analyzed.}
\label{fig4}
\end{center}
\end{figure}

Let us consider the mechanical system with two degrees of freedom
shown in Fig.\ref{fig4}, consisting of three rigid rods connected by
two elastic hinges of rotational rigidity $k$, and constrained at
one end by a hinge and at the other by a roller support. A mass $m$
is placed in correspondence of the intermediate elastic hinges and
the system is loaded by a horizontal axial force $N$. Assuming the
vertical displacements $x_1$ and $x_2$ of the elastic hinges as the
generalized coordinates, the total potential energy, $W$, and the
kinetic energy, $T$, of the whole system are given by:
\begin{equation}
\begin{aligned}\label{eq13}
W(x_1,x_2)=&\dfrac{1}{2}k\left[\left(\arcsin\dfrac{x_1}{l}-\arcsin\dfrac{x_2-x_1}{l}\right)^2\right.\\
&+\left.\left(\arcsin\dfrac{x_2}{l}+\arcsin\dfrac{x_2-x_1}{l}\right)^2\right]\\
&-Nl\left[3-\cos\left(\arcsin\dfrac{x_1}{l}\right)-\cos\left(\arcsin\dfrac{x_2}{l}\right)\right.\\
&-\left.\cos\left(\arcsin\dfrac{x_2-x_1}{l}\right)\right],\\
T(\dot{x_1},\dot{x_2})=&\dfrac{1}{2}m\dot{x_1}^2+\dfrac{1}{2}m\dot{x_1}^2
x_1^2+\dfrac{1}{2}m\dot{x_2}^2\\
&+\dfrac{1}{2}m\left(\dfrac{2x_1\dot{x_1}}{l}+\dfrac{x_2\dot{x_2}}{l}-\dfrac{x_2\dot{x_1}}{l}-\dfrac{x_1\dot{x_2}}{l}\right)^2.
\end{aligned}
\end{equation}

Performing a Taylor series expansion of Eq.\eqref{eq13} about the
origin, and assuming $x_1/l<1/10$ and $x_2/l<1/10$, we obtain:
\begin{equation}
\begin{aligned}\label{eq14}
W(x_1,x_2)&\cong \dfrac{k}{2l^2}\left(5x_1^2+5x_2^2-8x_1
x_2\right)-\dfrac{N}{l}\left(x_1^2+x_2^2-x_1 x_2\right),\\
T(\dot{x_1},\dot{x_2})& \cong
\dfrac{1}{2}m\dot{x_1}^2+\dfrac{1}{2}m\dot{x_2}^2.
\end{aligned}
\end{equation}

The equations of motion are identified by considering the Lagrange's
equations:
\begin{equation}\label{eq15}
\dfrac{\partial}{\partial t} \left(\dfrac{\partial
T}{\partial\dot{x_i}}\right)-\dfrac{\partial T}{\partial
x_i}=-\dfrac{\partial W}{\partial x_i},\qquad i=1,2.
\end{equation}
In matrix form, they are:
\begin{equation}\label{eq16}
\left[
\begin{array}{cc}
m & 0 \\
0 & m \\
\end{array}
\right]
\left\{
  \begin{array}{c}
    \ddot{x}_1 \\
    \ddot{x}_2 \\
  \end{array}
\right\}
+
\left[
\begin{array}{cc}
\dfrac{5k}{l^2} & -\dfrac{4k}{l^2} \\
-\dfrac{4k}{l^2} & \dfrac{5k}{l^2} \\
\end{array}
\right]
\left\{
  \begin{array}{c}
    x_1 \\
    x_2 \\
  \end{array}
\right\}
-N
\left[
\begin{array}{cc}
\dfrac{2}{l} & -\dfrac{1}{l} \\
-\dfrac{1}{l} & \dfrac{2}{l} \\
\end{array}
\right] \left\{
  \begin{array}{c}
    x_1 \\
    x_2 \\
  \end{array}
\right\}
=
\left\{
  \begin{array}{c}
   0 \\
   0 \\
  \end{array}
\right\}
\end{equation}

Looking for the solution to Eq.\eqref{eq16} in the general form
$\{q\}=\{q_0\}\mbox{e}^{\mbox{i} \omega t}$, where $\omega$ denotes
the natural angular frequency of the system, we obtain the following
equation, written in symbolic form:
\begin{equation}\label{eq17}
\left(-\omega^2[M]+[K]-N[K_g]\right)\{q_0\}=\{0\},
\end{equation}
where $[M]$, $[K]$ and $[K_G]$ denote, respectively, the mass
matrix, the elastic stiffness matrix and the geometric stiffness
matrix of the mechanical system. Their expressions can be simply
obtained by comparying Eq.\eqref{eq17} with Eq.\eqref{eq16}.

A nontrivial solution to Eq.\eqref{eq17} exists if and only if the
determinant of the resultant coefficient matrix of the vector
$\{q_0\}$ vanishes. This yields the following generalized eigenvalue
problem:
\begin{equation}\label{eq18}
\det\left([K]-N[K_g]-\omega^2[M]\right)=0,
\end{equation}
where $N$ and $\omega^2$ represent the eigenvalues. For this
example, Eq.\eqref{eq18} provides the following relationships
between the eigenvalues $\omega$ and $N$:
\begin{subequations}
\begin{align}
\omega^2 &= \dfrac{k}{ml^2}-\dfrac{N}{ml},\label{eq19a}\\
\omega^2 &= 3\dfrac{k}{ml^2}-\dfrac{3N}{ml}.\label{eq19b}
\end{align}
\end{subequations}
As limit cases, if $m=0$, then we obtain the Eulerian buckling
loads:
\begin{subequations}
\begin{align}\label{eq20}
N_1&=\dfrac{k}{l},\\
N_2&=\dfrac{3k}{l},
\end{align}
\end{subequations}
whereas, if $N=0$, we obtain the natural frequencies of the system:
\begin{subequations}
\begin{align}
\omega_1&=\sqrt{\dfrac{k}{ml^2}},\label{eq21a}\\
\omega_2&=\sqrt{\dfrac{9k}{ml^2}}.\label{eq21b}
\end{align}
\end{subequations}

As far as the eigenvectors are concerned, the system \eqref{eq17}
yields the eigenvectors corresponding, respectively, to the
eigenfrequencies \eqref{eq19a} and \eqref{eq19b} as functions of
$N$:
\begin{subequations}
\begin{align}\label{eq22}
x_1 &=
\dfrac{4k/l-N}{6k/l-3N}x_2,\\
x_1 &= \dfrac{4k/l-N}{14k/l-5N}x_2.
\end{align}
\end{subequations}

\begin{figure}
\begin{center}
\includegraphics*[width=.6\textwidth,angle=-90]{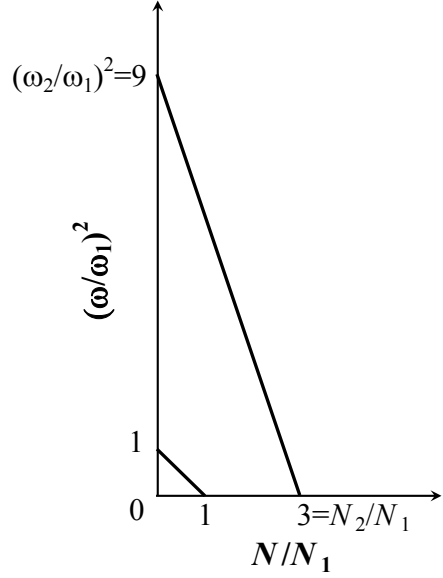}
\caption{Nondimensional frequencies vs. nondimensional axial forces
for the two-degrees of freedom system in Fig.\ref{fig4}.}
\label{fig5}
\end{center}
\end{figure}

Dividing Eqs.\eqref{eq19a} and \eqref{eq19b} by $\omega_1^2$, we
derive the following nondimensional relationships between the
eigenvalues:
\begin{subequations}
\begin{align}
\left(\dfrac{\omega}{\omega_1}\right)^2 &=1-\left(\dfrac{N}{N_1}\right),\label{eq23a}\\
\left(\dfrac{\omega}{\omega_1}\right)^2
&=\left(\dfrac{\omega_2}{\omega_1}\right)^2-\dfrac{N_2}{N_1}\left(\dfrac{N}{N_1}\right).\label{eq23b}
\end{align}
\end{subequations}
In analogy with the results for the single degree of freedom
systems, a graphical representation of Eqs.\eqref{eq23a} and
\eqref{eq23b} is provided in Fig.\ref{fig5}. We notice that both the
eigenfrequencies are decreasing functions of the compressive axial
load. Starting from $N=0$, bifurcation of the equilibrium would
correspond to pure resonance instability. Entering the diagram with
a value of the nondimensional compressive axial force in the range
$0<N/N_1<1$, the coordinates of the points of the two curves provide
the critical eigenfrequencies of the mechanical system leading to
bifurcation. Axial forces higher than $N_1$ in the range
$1<N/N_1<N_2/N_1$ can only be experienced if an additional
constraint is introduced into the system. Moreover, we observe that
the applied compressive load influences all the eigenfrequencies. In
particular, for the present example, the influence of the axial load
is greater on the highest frequency than on the lower one.

\begin{figure}
\begin{center}
\includegraphics*[width=.3\textwidth,angle=-90]{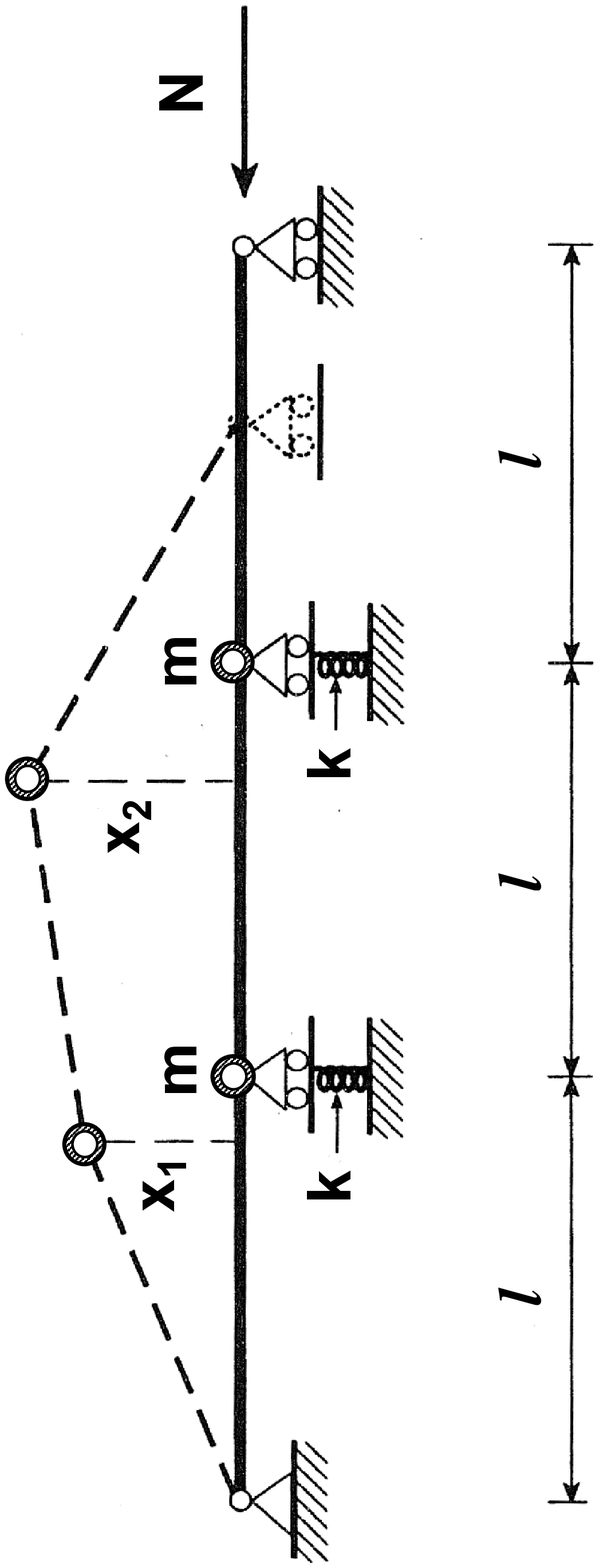}
\caption{Scheme of the second two-degrees of freedom system
analyzed.} \label{fig6}
\end{center}
\end{figure}

As a second example of a system with two degrees of freedom, let us
examine that of Fig.\ref{fig6}, which consists of three rigid rods
on four supports, of which the central ones are assumed to be
elastically compliant with rigidity $k$. A mass $m$ is placed in
correspondence of the intermediate hinges and the system is loaded
by a horizontal axial force $N$. Assuming the vertical displacements
$x_1$ and $x_2$ of the elastic hinges as the generalized
coordinates, the total potential energy, $W$, and the kinetic
energy, $T$, of the whole system are given by ($x_1/l<1/10$ and
$x_2/l<1/10$):
\begin{equation}
\begin{aligned}\label{eq24}
W(x_1,x_2)=&\dfrac{1}{2}k\left(x_1^2+x_2^2\right)-Nl\left[3-\cos\left(\arcsin\dfrac{x_1}{l}\right)\right.\\
&-\cos\left(\arcsin\dfrac{x_2}{l}\right)-\left.\cos\left(\arcsin\dfrac{x_2-x_1}{l}\right)\right]\\
\cong &
\dfrac{1}{2}k\left(x_1^2+x_2^2\right)-\dfrac{N}{l}\left(x_1^2+x_2^2-x_1
x_2\right),\\
T(\dot{x_1},\dot{x_2})\cong
&\dfrac{1}{2}m\dot{x_1}^2+\dfrac{1}{2}m\dot{x_1}^2.
\end{aligned}
\end{equation}

In this case, the Lagrange's equations \eqref{eq15} yield the
following matrix form:
\begin{equation}\label{eq25}
\left[
\begin{array}{cc}
m & 0 \\
0 & m \\
\end{array}
\right] \left\{
  \begin{array}{c}
    \ddot{x}_1 \\
    \ddot{x}_2 \\
  \end{array}
\right\} + \left[
\begin{array}{cc}
k & 0 \\
0 & k \\
\end{array}
\right] \left\{
  \begin{array}{c}
    x_1 \\
    x_2 \\
  \end{array}
\right\} -N \left[
\begin{array}{cc}
\dfrac{2}{l} & -\dfrac{1}{l} \\
-\dfrac{1}{l} & \dfrac{2}{l} \\
\end{array}
\right] \left\{
  \begin{array}{c}
    x_1 \\
    x_2 \\
  \end{array}
\right\} = \left\{
  \begin{array}{c}
   0 \\
   0 \\
  \end{array}
\right\}.
\end{equation}

Looking for the solution to Eq.\eqref{eq25} in the general form
$\{q\}=\{q_0\}\mbox{e}^{\mbox{i} \omega t}$, where $\omega$ denotes
the natural angular frequency of the system, we obtain the following
equation, written in symbolic form:
\begin{equation}\label{eq26}
\left(-\omega^2[M]+[K]-N[K_g]\right)\{q_0\}=\{0\},
\end{equation}
where $[M]$, $[K]$ and $[K_g]$ denote, respectively, the mass
matrix, the elastic stiffness matrix and the geometric stiffness
matrix of the mechanical system. As it can be readily seen, the
geometric stiffness matrix for this problem is the same as that of
the previous example.

A nontrivial solution to Eq.\eqref{eq26} exists if and only if the
determinant of the resultant coefficient matrix of the vector
$\{q_0\}$ is equal to zero. This yields the following generalized
eigenvalue problem:
\begin{equation}\label{eq27}
\det\left([K]-N[K_g]-\omega^2[M]\right)=0,
\end{equation}
where $N$ and $\omega^2$ are the eigenvalues of the system. For this
example, Eq.\eqref{eq27} provides the following relationships
between the eigenvalues:
\begin{subequations}
\begin{align}
\omega^2 &= \dfrac{k}{m}-3\dfrac{N}{ml},\label{eq28a}\\
\omega^2 &= \dfrac{k}{m}-\dfrac{N}{ml}.\label{eq28b}
\end{align}
\end{subequations}
As limit cases, if $m=0$, we obtain the Eulerian buckling loads:
\begin{subequations}
\begin{align}\label{eq29}
N_1&=\dfrac{1}{3}kl,\\
N_2&=kl,
\end{align}
\end{subequations}
whereas, if $N=0$, then we obtain the natural
frequencies of the system:
\begin{equation}\label{eq30}
\omega_1=\omega_2=\sqrt{\dfrac{k}{m}}.
\end{equation}

As far as the eigenvectors are concerned, the system \eqref{eq26}
yields the eigenvectors corresponding, respectively, to the
eigenfrequencies \eqref{eq28a} and \eqref{eq28b}, as functions of
the axial force, $N$:
\begin{subequations}
\begin{align}\label{eq31}
x_1 &=
\dfrac{N/l}{5N/l-2k}x_2,\\
x_1 &= \dfrac{N/l}{3N/l-2k}x_2.
\end{align}
\end{subequations}

Dividing Eqs.\eqref{eq28a} and \eqref{eq28b} by $\omega_1^2$, we
obtain the following nondimensional relationships between the
eigenvalues:
\begin{subequations}
\begin{align}
\left(\dfrac{\omega}{\omega_1}\right)^2 &=1-\left(\dfrac{N}{N_1}\right),\label{eq32a}\\
\left(\dfrac{\omega}{\omega_1}\right)^2
&=1-\dfrac{N_1}{N_2}\left(\dfrac{N}{N_1}\right).\label{eq32b}
\end{align}
\end{subequations}

\begin{figure}
\begin{center}
\includegraphics*[width=.35\textwidth,angle=-90]{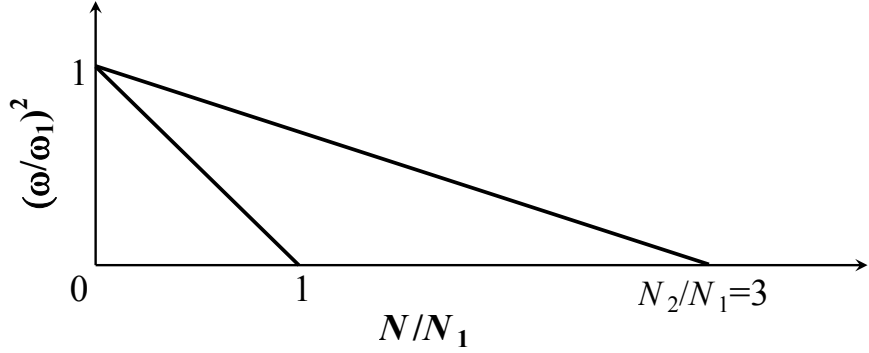}
\caption{Nondimensional frequencies vs. nondimensional axial forces
for the two-degrees of freedom system in Fig.\ref{fig6}.}
\label{fig7}
\end{center}
\end{figure}

A graphical representation of Eqs.\eqref{eq32a} and \eqref{eq32b} is
provided in Fig.\ref{fig7}. Also in this case, both the frequencies
are decreasing functions of the compressive axial load. However, for
the present example, the influence of the axial load is greater on
the lower frequency of the system than on the higher one.

\section{Continuous mechanical systems}

\subsection{Oscillations of deflected beams under compressive axial loads}

\begin{figure}
\begin{center}
\includegraphics*[width=.4\textwidth,angle=-90]{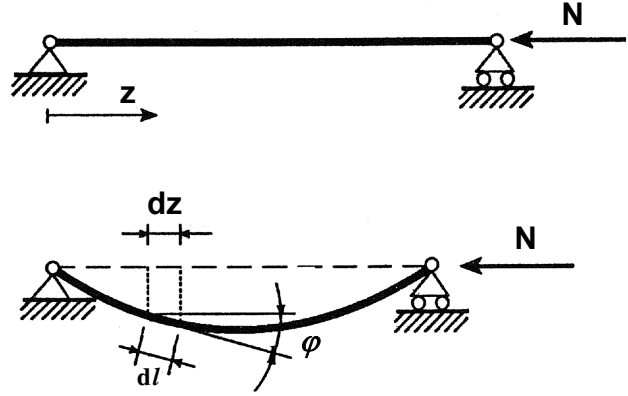}
\caption{Undeformed and deformed configurations of a deflected beam
under compressive axial force.} \label{fig8}
\end{center}
\end{figure}

Let us consider a slender elastic beam of constant cross-section,
inextensible and not deformable in shear, though deformable in
bending, constrained at one end by a hinge and at the other by a
roller support, loaded by an axial force, $N$ (see Fig.\ref{fig8}).
In this case, with the purpose of analyzing the free flexural
oscillations of the beam, the differential equation of the elastic
line with second-order effects can be written by replacing the
distributed load with the force of inertia:
\begin{equation}\label{eq33}
EI\dfrac{\partial^4 v}{\partial z^4}+N\dfrac{\partial^2 v}{\partial
z^2}=-\mu\dfrac{\partial^2 v}{\partial t^2},
\end{equation}
where $EI$ denotes the flexural rigidity of the beam and $\mu$ is
its linear density (mass per unit length). Equation \eqref{eq33} can
be rewritten in the following form:
\begin{equation}\label{eq34}
\dfrac{\partial^4 v}{\partial z^4}+\beta^2\dfrac{\partial^2
v}{\partial z^2}=-\dfrac{\mu}{EI}\dfrac{\partial^2 v}{\partial t^2},
\end{equation}
where we have set $\beta^2=N/EI$.

Equation \eqref{eq34} is an equation with separable variables, the
solution being represented as the product of two different
functions, each one depending on a single variable:
\begin{equation}\label{eq35}
v(z,t)=\eta(z)f(t).
\end{equation}
Introducing Eq.\eqref{eq35} into Eq.\eqref{eq34}, leads:
\begin{equation}\label{eq36}
\dfrac{\mbox{d}^4 \eta}{\mbox{d} z^4}f+\beta^2\dfrac{\mbox{d}^2
\eta}{\mbox{d} z^2}f+\dfrac{\mu}{EI}\eta\dfrac{\mbox{d}^2
f}{\mbox{d} t^2}=0.
\end{equation}
Dividing Eq.\eqref{eq36} by the product $\eta f$, we find:
\begin{equation}\label{eq37}
-\dfrac{\dfrac{\mbox{d}^2
f}{\mbox{d}t^2}}{f}=\dfrac{\mu}{EI}\dfrac{\dfrac{\mbox{d}^4
\eta}{\mbox{d}
z^4}+\beta^2\dfrac{\mbox{d}^2\eta}{\mbox{d}z^2}}{\eta}=\omega^2,
\end{equation}
where $\omega^2$ represents a positive constant, the left and the
right hand-sides of Eq.\eqref{eq37} being at the most functions of
the time $t$ and the coordinate $z$, respectively. From
Eq.\eqref{eq37} there follow two ordinary differential equations:
\begin{subequations}
\begin{align}
\dfrac{\mbox{d}^2 f}{\mbox{d} t^2}+\omega^2
f &=0,\label{eq38a}\\
\dfrac{\mbox{d}^4 \eta}{\mbox{d}
z^4}+\beta^2\dfrac{\mbox{d}^2\eta}{\mbox{d}z^2}-\alpha^4\eta
&=0,\label{eq38b}
\end{align}
\end{subequations}
with
\begin{equation}\label{eq39}
\alpha=\sqrt[4]{\dfrac{\mu \omega^2}{EI}}.
\end{equation}
Whereas Eq.\eqref{eq38a} is the equation of the harmonic oscillator,
with the well-known complete integral
\begin{equation}\label{eq40}
f(t)=A\cos\omega t+B\sin\omega t,
\end{equation}
Eq.\eqref{eq38b} has the following complete integral
\begin{equation}\label{eq41}
\eta(z)=C\mbox{e}^{\lambda_1 z}+D\mbox{e}^{\lambda_2
z}+E\mbox{e}^{-\lambda_1 z}+F\mbox{e}^{-\lambda_2 z},
\end{equation}
where $\lambda_1$ and $\lambda_2$ are functions of $\alpha$ and
$\beta$:
\begin{equation}\label{eq42}
\lambda_{1,2}=\sqrt{\dfrac{-\beta^2\pm
\sqrt{\beta^4+4\alpha^4}}{2}}.
\end{equation}
As in the modal analysis, the constants $A$ and $B$ can be
determined on the basis of the initial conditions, while the
constants $C$, $D$, $E$ and $F$ can be determined by imposing the
boundary conditions. As it will be shown in the sequel, for a given
value of $\beta$, the parameters $\omega$ and $\alpha$ can be
determined by solving a generalized eigenvalue problem resulting
from the imposition of the boundary conditions. From the
mathematical point of view, this eigenvalue problem is analogous to
that shown for the discrete systems. On the other hand, since we are
considering a continuous mechanical system having infinite degrees
of freedom, we shall obtain an infinite number of eigenvalues
$\omega_i$ and $\alpha_i$, just as also an infinite number of
eigenfunctions $f_i$ and $\eta_i$. The complete integral of the
differential equation \eqref{eq33} may therefore be given the
following form, according to the Principle of Superposition:
\begin{equation}\label{eq43}
v(z,t)=\sum_{i=1}^{\infty}\eta_i(z)f_i(t),
\end{equation}
with:
\begin{subequations}
\begin{align}
f_i(t)&= A_i\cos\omega_i t+B_i\sin\omega_i t,\label{eq44a}\\
\eta(z)&= C_i\mbox{e}^{\lambda_{1i} z}+D_i\mbox{e}^{\lambda_{2i}
z}+E_i\mbox{e}^{-\lambda_{1i} z}+F_i\mbox{e}^{-\lambda_{2i}
z}.\label{eq44b}
\end{align}
\end{subequations}
It is important to remark that the eigenfunctions $\eta_i$ are still
orthonormal functions, as in the classical modal analysis (see the
mathematical demonstration reported in the Appendix). This permits
to determine the constants $A_i$ and $B_i$ in Eq.\eqref{eq44a} via
the initial conditions (see also \cite{aC97}, Pag. 315):
\begin{subequations}
\begin{align}
v(z=0)&=v_0(z),\label{eq45a}\\
\dfrac{\partial v}{\partial t}(z=0)&=\dot{v}_0(z).\label{eq45b}
\end{align}
\end{subequations}

As regards the boundary conditions, let us consider as an example a
beam supported at both ends, of length $l$:
\begin{equation}\label{eq46}
\left\{
  \begin{array}{ll}
    \eta(0) &=0,\\
    \eta^{''}(0) &=0,\\
    \eta(l) &=0,\\
    \eta^{''}(l) &=0,
  \end{array}
\right.
\Rightarrow
\left[
  \begin{array}{cccc}
    1 & 1 & 1 & 1\\
    \lambda_1^2 & \lambda_2^2 & \lambda_1^2 & \lambda_2^2\\
    \mbox{e}^{\lambda_1 l} & \mbox{e}^{\lambda_2 l} & \mbox{e}^{-\lambda_1 l} & \mbox{e}^{-\lambda_2 l}\\
    \lambda_1^2\mbox{e}^{\lambda_1 l} & \lambda_2^2\mbox{e}^{\lambda_2 l} & \lambda_1^2\mbox{e}^{-\lambda_1 l} & \lambda_2^2\mbox{e}^{-\lambda_2 l}
  \end{array}
\right]
\left\{
\begin{array}{c}
C\\
D\\
E\\
F
\end{array}
\right\}
=
\left\{
\begin{array}{c}
0\\
0\\
0\\
0
\end{array}
\right\},
\end{equation}

For a nontrivial solution to the system in Eq.\eqref{eq46}, the
determinant of the coefficient matrix has to vanish. The resulting
eigenequation permits, for each given value of the parameter
$\beta$, to determine the eigenvalues $\alpha_i$ of the system.
Finally, the corresponding natural eigenfrequencies $\omega_i$ can
be obtained by inverting Eq.\eqref{eq39}.

\begin{figure}
\begin{center}
\includegraphics*[width=.6\textwidth,angle=-90]{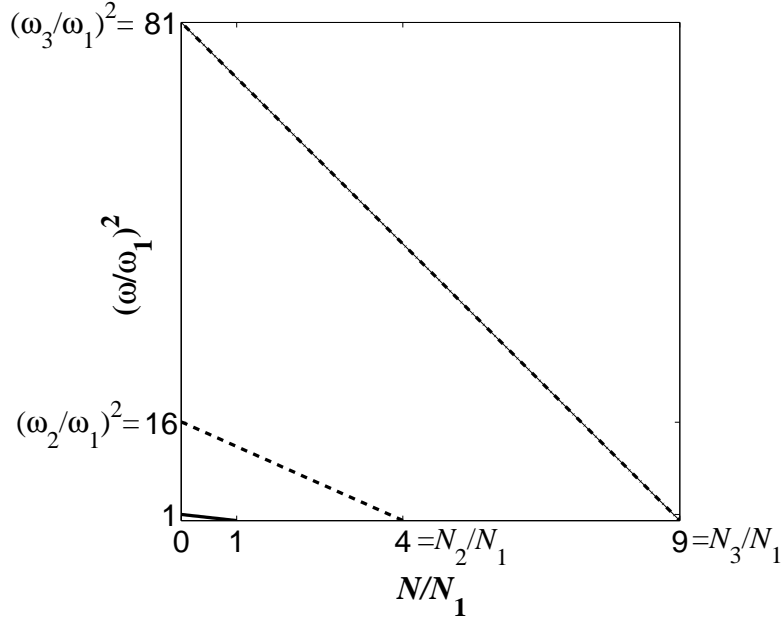}
\caption{Nondimensional frequencies vs. nondimensional axial force
for the continuous system in Fig.\ref{fig8}.} \label{fig9}
\end{center}
\end{figure}

As an illustrative example, the first three nondimensional
frequencies of the simply supported beam shown in Fig.\ref{fig8} are
reported in Fig.\ref{fig9} as functions of the applied
nondimensional axial force. Parameters $\omega_i$ and $N_i$ denote,
respectively, the $i$-th frequency of the system determined
according to modal analysis and the $i$-th buckling load determined
according to the Euler's formula. In close analogy with the discrete
mechanical systems, the curves in the $(\omega/\omega_1)^2$ vs.
$N/N_1$ plane are represented by straight lines. Also in this case,
the coordinates of the points along these lines provide the critical
conditions leading to the system instability in terms of frequency
of the excitation and magnitude of the applied compressive axial
force.

\subsection{Oscillations and lateral-torsional buckling of beams}

\begin{figure}
\begin{center}
\includegraphics*[width=.7\textwidth,angle=180]{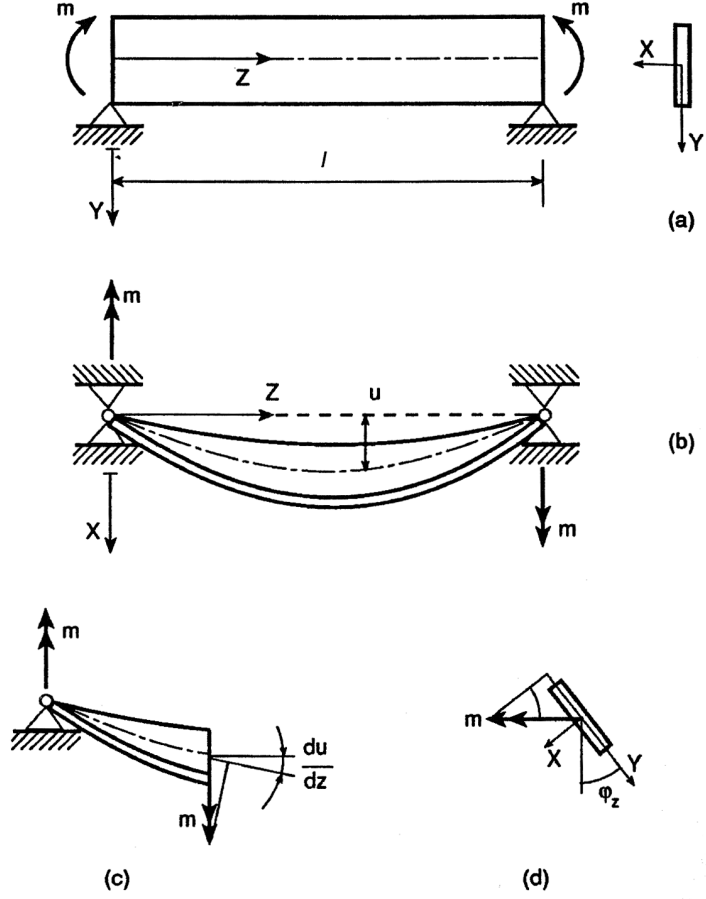}
\caption{Scheme of a beam subjected to lateral-torsional buckling.}
\label{fig10}
\end{center}
\end{figure}

Let us consider a beam of thin rectangular cross-section,
constrained at the ends so that rotation about the longitudinal axis
$Z$ is prevented. Let this beam be subjected to uniform bending by
means of the application at the ends of two moments  $m$ contained
in the plane $YZ$ of greater flexural rigidity (see
Fig.\ref{fig10}).

Considering a deformed configuration of the beam, with deflection
thereof in the $XZ$ plane of smaller flexural rigidity, and
simultaneous torsion about the $Z$ axis (see Fig.\ref{fig10}),
bending-torsional out-of-plane vibrations of the beam are described
by the following partial differential equations:
\begin{equation}\label{eq47}
\begin{aligned}
EI_y\dfrac{\partial^4 u}{\partial z^4}+m\dfrac{\partial^2
\varphi_z}{\partial z^2}=&-\mu\dfrac{\partial^2 u}{\partial t^2},\\
-GI_t\dfrac{\partial^2 \varphi_z}{\partial z^2}+m\dfrac{\partial^2
u}{\partial z^2}=&-\mu\rho^2\dfrac{\partial^2 \varphi_z}{\partial
t^2},
\end{aligned}
\end{equation}
where $u(z,t)$ and $\varphi_z(z,t)$ are, respectively, the
out-of-plane deflection and the twist angle of the beam
cross-section; $EI_y$ and $GI_t$ are the bending and torsional
rigidities; $\mu$ is the mass of the beam per unit length, and
$\rho=\sqrt{I_P/A}$ is the polar radius of inertia of the beam
cross-section.

A solution to the system \eqref{eq47} can be found in the following
variable-separable form \cite{vvB95}:
\begin{equation}\label{eq48}
\begin{aligned}
u(z,t)=&U(t)\eta(z),\\
\varphi_z(z,t)=&\Phi(t)\psi(z),
\end{aligned}
\end{equation}
where the functions $\eta(z)$ and $\psi(z)$ are such that the
boundary conditions
$\eta(0)=\eta(l)=\eta^{''}(0)=\eta^{''}(l)=\psi(0)=\psi(l)=0$ are
satisfied. According to Bolotin \cite{vvB95}, we can assume
$\eta(z)=\psi(z)=\sin\dfrac{n\pi z}{l}$, with $n$ being a natural
number. In this case, we obtain the following matrix form:
\begin{equation}\label{eq49}
\left[
\begin{array}{cc}
\mu & 0 \\
0 & \mu\rho^2 \\
\end{array}
\right] \left\{
  \begin{array}{c}
    \ddot{U} \\
    \ddot{\Phi} \\
  \end{array}
\right\} + \left[
\begin{array}{cc}
EI_y\dfrac{n^4\pi^4}{l^4} & 0 \\
0 & GI_t\dfrac{n^2\pi^2}{l^2} \\
\end{array}
\right] \left\{
  \begin{array}{c}
    U \\
    \Phi \\
  \end{array}
\right\} -m \left[
\begin{array}{cc}
0 & \dfrac{n^2\pi^2}{l^2} \\
\dfrac{n^2\pi^2}{l^2} & 0 \\
\end{array}
\right] \left\{
  \begin{array}{c}
    U \\
    \Phi \\
  \end{array}
\right\} = \left\{
  \begin{array}{c}
   0 \\
   0 \\
  \end{array}
\right\},
\end{equation}
which can be symbolically rewritten as:
\begin{equation}\label{eq50}
\left[M\right]\{\ddot{q}\}+\left[K\right]\{q\}-m\left[K_g\right]\{q\}=\{0\},
\end{equation}
where $\left\{q\right\}=(U,\Phi)^{\textsf{T}}$. The mass matrix,
$\left[ M \right]$, the elastic stiffness matrix $\left[K\right]$,
and the geometric stiffness matrix $\left[K_g\right]$ in
Eq.\eqref{eq50} can be defined in comparison with Eq.\eqref{eq49}.
Looking for a general solution in the form
$\{q\}=\{q_0\}\mbox{e}^{\mbox{i} \omega t}$, we obtain:
\begin{equation}\label{eq51}
\left(\left[K\right]-m\left[K_g\right]-\omega^2\left[M\right]\right)\{q_0\}=\{0\}.
\end{equation}
A nontrivial solution to Eq.\eqref{eq51} exists if and only if the
determinant of the resultant coefficient matrix of the vector
$\{q_0\}$ vanishes. This yields the following generalized eigenvalue
problem:
\begin{equation}\label{eq52}
\det\left(\left[K\right]-m\left[K_g\right]-\omega^2\left[M\right]\right)=0,
\end{equation}
where $m$ and $\omega^2$ are the eigenvalues of the system.

As limit cases, if $\mu=0$, then we obtain the critical bending
moments given by the Prandtl's formula:
\begin{equation}\label{eq53}
m_{nc}=\dfrac{n\pi}{l}\sqrt{EI_y GI_t},
\end{equation} whereas, if $m=0$, then we obtain the natural flexural and torsional eigenfrequencies of the beam:
\begin{equation}\label{eq54}
\begin{aligned}
\omega^{\text{flex}}_{n}=& \left(\dfrac{n\pi}{l}\right)^2\sqrt{\dfrac{EI_y}{\mu}},\\
\omega^{\text{tors}}_{n}=& \dfrac{n\pi}{\rho
l}\sqrt{\dfrac{GI_t}{\mu}}.
\end{aligned}
\end{equation}

\begin{figure}
\begin{center}
\includegraphics*[width=.65\textwidth,angle=-90]{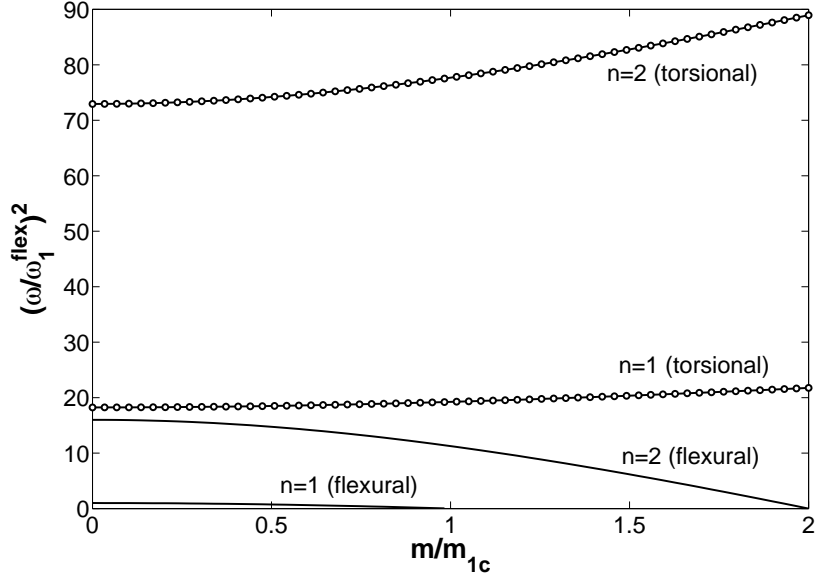}
\caption{Nondimensional flexural and torsional frequencies vs.
nondimensional bending moments for the continuous system in
Fig.\ref{fig10}.} \label{fig11}
\end{center}
\end{figure}

Considering a rectangular beam with a depth to span ratio of $1/3$
and with a thickness to depth ratio of $1/10$, the evolution of the
first two flexural and torsional eigenfrequencies of the system are
shown in Fig.\ref{fig11} as functions of the applied bending moment.
In this case, the curves in the nondimensional plane
$(\omega/\omega^{\text{flex}}_1)^2$ vs. $m/m_{1c}$ are no longer
straight lines. This fact can be ascribed to the coupling between
torsional and flexural vibrations of the beam. Moreover, when $m$ is
increased from zero (pure resonance instability) up to the critical
bending moment computed according to the Parandtl's formula (pure
buckling instability), $m_{1c}$, we note that the resonance
frequencies related to flexural oscillations progressively decrease
from $\omega^{\text{flex}}_1$ down to zero in correspondence of the
critical bending moments given by the Prandtl's formula. Conversely,
the resonance frequencies related to torsional oscillations
increase. From the mathematical point of view, this is the result of
the fact that the sum of the squares of the two eigenfrequencies for
a given value of $n$ is constant when the applied moment $m$ is
varied.

\section{Finite elements procedure}

When the mechanical system cannot be reduced to the schemes
previously analyzed, it is possible to apply the Finite Element
Method \cite{kjB82,ZT05}. According to this approach, the equations
of motion for an elastic system with a finite number of degrees of
freedom can be expressed in matrix form, taking also into account
the effect of the geometric nonlinearity through the geometric
stiffness matrix \cite{aR82,jsP85}.

For the sake of generality, let the elastic domain $V$ be divided
into subdomains $V_e$, and let each element contain $m$ nodal
points, each one having $g$ degrees of freedom. In compact form, the
displacement vector field defined on the element $V_e$ may be
represented as:
\begin{equation}\label{eq55}
\begin{Large}
\begin{subarray}{c}
\left\{ \eta_e \right\}
\\
(g\times 1)
\end{subarray}
\begin{subarray}{c}
\,=\\
«
\end{subarray}
\begin{subarray}{c}
\left[\eta_e\right]
\\
g\times(g\times m)
\end{subarray}
\,
\begin{subarray}{c}
\left\{\delta_e\right\}
\\
(g\times m)\times 1
\end{subarray}
\begin{subarray}{c}
\, ,\\
«
\end{subarray}
\end{Large}
\end{equation}
where $\left[\eta_e\right]$ is the matrix collecting the shape
functions and $\left\{\delta_e\right\}$ is the nodal displacements
vector.

The deformation characteristic vector is obtained by derivation:
\begin{equation}\label{eq56}
\begin{Large}
\begin{subarray}{c}
\left\{q_e\right\}
\\
(d\times 1)
\end{subarray}
\begin{subarray}{c}
\,=\\
«
\end{subarray}
\begin{subarray}{c}
\left[\partial\right]
\\
(d\times g)
\end{subarray}
\,
\begin{subarray}{c}
\left\{\eta_e\right\}
\\(g\times 1)
\end{subarray}
\begin{subarray}{c}
\, ,\\
«
\end{subarray}
\end{Large}
\end{equation}
where $\left[\partial\right]$ is the kinematic operator relating
strains to displacements, whereas $d$ denotes the dimension of the
strain characteristic vector, e.g. $d=3$ for a beam in plane or
$d=6$ for a beam in space. Hence, introducing Eq.\eqref{eq55} into
Eq.\eqref{eq56}, we obtain:
\begin{equation}\label{eq57}
\begin{Large}
\begin{subarray}{c}
\left\{q_e\right\}
\\
(d\times 1)
\end{subarray}
\begin{subarray}{c}
\,=\\
«
\end{subarray}
\begin{subarray}{c}
\left[\partial\right]
\\
(d\times g)
\end{subarray}
\;
\begin{subarray}{c}
\left[\eta_e\right]
\\
g\times(g\times m)
\end{subarray}
\,
\begin{subarray}{c}
\left\{\delta_e\right\}
\\
(g\times m)\times 1
\end{subarray}
\begin{subarray}{c}
\,=\\
«
\end{subarray}
\begin{subarray}{c}
\left[B_e\right]
\\
d\times (g\times m)
\end{subarray}
\,
\begin{subarray}{c}
\left\{\delta_e\right\}
\\
(g\times m)\times 1
\end{subarray}
\begin{subarray}{c}
\, ,\\
«
\end{subarray}
\end{Large}
\end{equation}
where the matrix $\left[B_e\right]$ is calculated by derivation of
the shape functions.

According to these definitions, we can obtain the following matrix
equation for each finite element (see \cite{aC97}, Ch.11 for more
details):
\begin{equation}\label{eq58}
\left[M_e\right]\left\{\ddot{\delta_e}\right\}+\left(\left[K_e\right]-\left[K_{ge}\right]\right)\left\{\delta_e\right\}=\left\{0\right\},
\end{equation}
where $\left[M_e\right]$, $\left[K_e\right]$ and
$\left[K_{ge}\right]$ denote, respectively, the local mass matrix,
the local elastic stiffness matrix and the local geometric stiffness
matrix of the finite element.

As usual, the local mass and elastic stiffness matrices are given by
\cite{kjB82,ZT05}:
\begin{subequations}
\begin{align}
\left[M_e\right]=&\int_{V_e}\left[\eta_e\right]^{\textsf{T}}\left[\mu\right]\left[\eta_e\right]\mbox{d}V,\label{eq58a}\\
\left[K_e\right]=&\int_{V_e}\left[B_e\right]^{\textsf{T}}\left[H\right]\left[B_e\right]\mbox{d}V.\label{eq58b}
\end{align}
\end{subequations}
The local geometric stiffness matrix can be computed as follows
\cite{aR82}:
\begin{equation}\label{eq59}
\left[K_{ge}\right]=\int_{V_e}\left[G_e\right]^{\textsf{T}}\left[S_e\right]\left[G_e\right]\mbox{d}V,
\end{equation}
where the matrix $\left[S_e\right]$ is related to the components of
the stress field:
\begin{equation}\label{eq60}
\left[S_e\right] = \left[
\begin{array}{ccc}
    \sigma_x\left[I_3\right] & \tau_{xy}\left[I_3\right]  & \tau_{xz}\left[I_3\right]\\
    \tau_{xy}\left[I_3\right]& \sigma_{y}\left[I_3\right] & \tau_{yz}\left[I_3\right]\\
    \tau_{xz}\left[I_3\right]& \tau_{yz}\left[I_3\right]  & \sigma_{z}\left[I_3\right]
  \end{array}
\right],
\end{equation}
where $\left[I_3\right]$ denotes a unit matrix with dimensions
$(3\times 3)$. The matrix $\left[G_e\right]$ is related to the first
derivative of the shape functions through the differential operator
$\left[\widehat{\partial}\right]$:
\begin{equation}\label{eq61}
\begin{Large}
\begin{subarray}{c}
\left[G_e\right]\\
(g\times g)\times (g\times m)
\end{subarray}
\begin{subarray}{c}
\,=\\
«
\end{subarray}
\begin{subarray}{c}
\left[\widehat{\partial}\right]\\
(g\times g)\times g
\end{subarray}
\;
\begin{subarray}{c}
[\eta_e]\\
g\times(g\times m)
\end{subarray}
\begin{subarray}{c}
\, ,\\
«
\end{subarray}
\end{Large}
\end{equation}
where
\begin{equation}\label{eq61b}
\left[\widehat{\partial}\right] = \left[
\begin{array}{c}
    \dfrac{\partial}{\partial x}\left[I_3\right] \\
    \dfrac{\partial}{\partial y}\left[I_3\right] \\
    \dfrac{\partial}{\partial z}\left[I_3\right]
  \end{array}
\right].
\end{equation}
According to this formulation, we note that the geometric stiffness
matrix is a function of the stress components through the matrix
$\left[S_e\right]$. In the case of a compressive stress field, the
geometric stiffness terms become negative and reduce the
corresponding elements of the local elastic stiffness matrix, just
as shown for the discrete mechanical systems. We also remark that
this formulation is quite general, since the information related to
the finite element topology is simply included in the matrix
$\left[\eta_e\right]$ which collects the shape functions and in the
differential operator $\left[\partial\right]$ (see \cite{aC97} for
more details).

By performing the usual operations of rotation, expansion and
assemblage of the mass, elastic stiffness and geometric stiffness
matrices of the element, Eq.\eqref{eq58} can be written in global
form:
\begin{equation}\label{eq62}
\left[M\right]\left\{\ddot{\delta}\right\}+\left(\left[K\right]-\lambda\left[K_{g}\right]\right)\left\{\delta\right\}=\left\{0\right\}.
\end{equation}
Looking for the solution to Eq.\eqref{eq62} in the general form
$\{\delta\}=\{\delta_0\}\mbox{e}^{\mbox{i} \omega t}$, where
$\omega$ is the natural frequency of the system, we can formulate
the generalized eigenproblem as in the cases discussed above:
\begin{equation}\label{eq63}
\det\left(\left[K_e\right]-\lambda\left[K_{g}\right]-\omega^2\left[M_e\right]\right)=0.
\end{equation}
Therefore, the numerical procedure for the determination of the
frequency-loading multiplier diagram consists in the following
steps.
\begin{enumerate}
\item For a given loading configuration defined by the loading multiplier $\lambda$, determine the stress field
according to a linear elastic stress analysis.
\item Compute the local mass matrix, the local elastic stiffness matrix and the local geometric stiffness matrix for each finite element.
\item Perform the rotation, expansion and assemblage operations to
obtain the global matrices.
\item Solve the generalized eigenvalue problem of Eq.\eqref{eq63} and find
the eigenfrequencies of the system, $\omega_i^2$, with
$i=1,\dots,g\times n$.
\item Iterate the above-described procedure for different values of
$\lambda$.
\end{enumerate}

\section{Discussion and conclusions}

The problems of elastic instability (buckling) and dynamic
instability (resonance) have been the subject of extensive
investigation and have received a large attention from the
structural mechanics community. Nonetheless, the study of the
interaction between these elementary forms of instability is still
an open point.

The phenomenon of flutter instability of the Tacoma Narrows Bridge
occurred on November 7, 1940, can be reinterpreted as the result of
such a catastrophic interaction. This cable-suspended bridge was
solidly built, with girders of carbon steel anchored in huge blocks
of concrete and was the first of its type to employ plate girders to
support the roadbed. While in the earlier designs any wind would
simply pass through the truss, in the new design of the 1940 the
wind would be diverted above and below the structure. Shortly after
construction, it was discovered that the bridge would sway and
buckle dangerously in windy conditions. This resonance was flexural,
meaning the bridge buckled along its length, with the roadbed
alternately raised and depressed in certain locations. However, the
failure of the bridge occurred when a never-before-seen twisting
mode occurred.

A Report to the Federal Works Agency \cite{AKW41} excluded the
phenomenon of pure forced resonance as the actual reason of
instability: "\emph{...it is very improbable that resonance with
alternating vortices plays an important role in the oscillations of
suspension bridges. First, it was found that there is no sharp
correlation between wind velocity and oscillation frequency such as
is required in case of resonance with vortices whose frequency
depends on the wind velocity...}". A new theory for the
interpretation of these complex aerodynamic instabilities was
developed by Scanlan \cite{ST71,rhS78} and then elaborated by
various researchers \cite{CGM85,NAB92,BD93,BB05}. Basically, the
so-called \emph{flutter theory} considers the following equation of
motion for the mechanical system in the finite element framework
\cite{NAB92,uS93}:
\begin{equation}
\left[M\right]\left\{\ddot{\delta}\right\}+\left[C\right]\left\{\dot{\delta}\right\}+\left[K\right]\left\{\delta\right\}=\left\{F\right\}_{mi}+\left\{F\right\}_{md},
\end{equation}
where $\left\{F\right\}_{mi}$ and $\left\{F\right\}_{md}$ are,
respectively, the motion-independent wind force vector and the
motion-dependent aeroelastic force vector. A special attention is
given to the structural damping, which is included in the equations
of motion through the damping matrix $[C]$. The motion-dependent
force vector is then put in relationship with the nodal
displacements of the system, $\{\delta\}$, and the nodal velocities,
$\{\dot{\delta}\}$, according to the \emph{flutter derivative
matrices}, $[K^*]$ and $[C^*]$, that are empirically determined in
the wind tunnel by using section models of the bridge. As a result,
the problem becomes highly nonlinear, and the flutter velocity,
$U_{cr}$, and the flutter frequency, $\omega_{cr}$, can be
determined from the following eigenproblem:
\begin{equation}
\det\left(\left[K\right]-\dfrac{1}{2}\rho
U_{cr}^2\left[K^*\right]-\omega^2_{cr}\left[M\right]+\omega_{cr}\left[C\right]-\dfrac{1}{2}\rho
U_{cr}\omega_{cr}\left[C^*\right]\right)=0,
\end{equation}
where $\rho$ is the air density and $1/2\rho U_{cr}^2$ is the wind
pressure.

It is important to remark that this eigenproblem shares most of the
features of the generalized eigenproblem that we have analyzed in
the present study. In fact, in both cases, two eigenvalues have to
be determined from the eigenequation. However, the flutter theory
gives prominence to the role played by the structural damping,
although being generally less than $1\%$ (see e.g.\cite{NAB92}).
Moreover, the value of the mechanical damping seems to represent a
sort of free parameter in the model. In fact, this parameter is
usually assumed, like in \cite{NAB92}, rather than experimentally
evaluated. Another difference with our proposed approach relies in
the geometric stiffness matrix, which is not taken into account in
the current flutter theory. However, due to the large structural
displacements, the flutter derivative matrix $[K^*]$ plays a very
similar role, although being experimentally obtained, rather than
analytically computed.

In conclusion, it seems to be possible to reinterpret the phenomenon
of aeroelastic instability as the result of the interaction between
pure resonance and pure buckling instabilities. According to our
approach, the geometric stiffness matrix has a preeminent role and
the mechanical damping can be neglected, as usually done in most of
the structural engineering applications. On this line, the collapse
of the Tacoma Narrows bridge can be considered as the result of the
interaction between buckling (related to the wind pressure
proportional to the square of the wind velocity) and resonance
(caused by the frequency of the wind gusts). Thus, this would give a
new explanation on why the Tacoma Narrows bridge failure took place
under moderate wind velocities (wind pressure lower than the
critical buckling load) and wind gusts frequencies different from
the natural frequencies of the bridge. Future developments of the
present work will regard the assessment of the proposed approach to
the analysis of bridge instabilities, as well as the comparison with
the classical flutter theory on the basis of real case histories.

\section{Appendix: orthonormality of the eigenfunctions of deflected beams subjected to an axial force}

As is well-known, the eigenfunctions $\eta_i$ of deflected beams
computed according to pure modal analysis are orthonormal functions.
It is possible to demonstrate that this fundamental property still
holds when the beam is subjected to an axial load, $N$, as that
shown in Fig.\ref{fig8}. We may in fact write Eq.\eqref{eq38b} for
two different eigensolutions:
\begin{subequations}
\begin{align}
\eta_j^{IV}+\beta^2\eta_j^{II}=\alpha_j^4\eta_j,\label{a1}\\
\eta_k^{IV}+\beta^2\eta_k^{II}=\alpha_k^4\eta_k.\label{a2}
\end{align}
\end{subequations}
Multiplying Eq.\eqref{a1} by $\eta_k$ and Eq.\eqref{a2} by $\eta_j$,
and integrating over the beam length, we obtain:
\begin{subequations}
\begin{align}
\int_0^l\eta_k\eta_j^{IV}\mbox{d}z+\beta^2\int_0^l\eta_k\eta_j^{II}\mbox{d}z=\alpha_j^4\int_0^l\eta_k\eta_j\mbox{d}z,\label{a3}\\
\int_0^l\eta_j\eta_k^{IV}\mbox{d}z+\beta^2\int_0^l\eta_j\eta_k^{II}\mbox{d}z=\alpha_k^4\int_0^l\eta_j\eta_k\mbox{d}z.\label{a4}
\end{align}
\end{subequations}
Integrating by parts the left-hand sides, the foregoing equations
transform as follows:
\begin{subequations}
\begin{align}
\begin{split}
&\left[\eta_k\eta_j^{III}\right]_0^l-\left[\eta_k^I\eta_j^{II}\right]_0^l+\int_0^l\eta_k^{II}\eta_j^{II}\mbox{d}z+\beta^2\left[\eta_k\eta_j\right]_0^l\\&-\beta^2\int_0^l\eta_k^I\eta_j^I\mbox{d}z=\alpha_j^4\int_0^l\eta_k\eta_j\mbox{d}z,
\end{split}\label{a5}\\
\begin{split}
&\left[\eta_j\eta_k^{III}\right]_0^l-\left[\eta_j^I\eta_k^{II}\right]_0^l+\int_0^l\eta_j^{II}\eta_k^{II}\mbox{d}z+\beta^2\left[\eta_j\eta_k\right]_0^l\\&-\beta^2\int_0^l\eta_j^I\eta_k^I\mbox{d}z=\alpha_k^4\int_0^l\eta_j\eta_k\mbox{d}z.\label{a6}
\end{split}
\end{align}
\end{subequations}
When each of the two ends of the beam is constrained by a built-in
support $(\eta=\eta^I=0)$, or by a hinge $(\eta=\eta^{II}=0)$, the
quantities in square brackets vanish. On the other hand, when the
end in $z=0$ is either unconstrained $(\eta^{III}=\eta^{II}=0)$, or
constrained by a double rod $(\eta^{III}=\eta^{I}=0)$, the remaining
end of the beam has to be constrained either by a built-in support
$(\eta=\eta^I=0)$, or by a simple support $(\eta=\eta^{II}=0)$. For
both configurations, only the terms $\left[\eta_i\eta_k\right]_0^l$
are different from zero.

In any case, subtracting member by member, these quantities are
canceled and we have:
\begin{equation}
\left(\alpha_j^4-\alpha_k^4\right)\int_0^l \eta_j\eta_k=0,
\end{equation}
which leads to the orthonormality condition:
\begin{equation}
\int_0^l \eta_j\eta_k=\delta_{ij},
\end{equation}
where $\delta_{ij}$ is the Kronecker delta. Thus, when the
eigenvalues are distinct, the integral of the product of the
corresponding eigenfunctions vanishes. When, instead, the indices
$j$ and $k$ coincide, the condition of normality reminds us that the
eigenfunctions are defined neglecting a factor of proportionality.

\vspace{1em} \noindent\textbf{Acknowledgements} \vspace{0.5em}

\noindent The financial support provided by the European Union to
the Leonardo da Vinci project “Innovative Learning and Training on
Fracture (ILTOF)” is gratefully acknowledged.

%\bibliographystyle{unsrt}
%\bibliography{biblio}

\end{document}